# Magnetization Current Simulation of High Temperature Bulk Superconductors Using *A-V-A* Formulation and Iterative Algorithm Method: Critical State Model and Flux Creep Model

Kai Zhang, Sebastian Hellmann, Marco Calvi [*], Thomas Schmidt

*Paul Scherrer Institute, Villigen, CH, 5232*

Lucas Brouwer

*Lawrence Berkeley National Laboratory, Berkeley, CA, 94720*

*\* marco.calvi@psi.ch*

## Abstract

In this work we will introduce the *A-V-A* formulation based iterative algorithm method (IAM) for simulating the magnetization current of high temperature superconductors. This new method embedded in ANSYS can simulate the critical state model by forcing the trapped current density to the critical current density $J_c$ for all meshed superconducting elements after each iterative load step, as well as simulate the flux creep model by updating the *E-J* power law based resistivity values. The simulation results of a disk-shaped ReBCO bulk during zero field cooling (ZFC) or field cooling (FC) magnetization agree well with the simulation results from using the *H*-formulation in COMSOL. The computation time is shortened by using the *A-V* formulation in superconductor areas and the *A*-formulation in non-superconductor areas. This iterative method is further proved friendly for adding ferromagnetic materials into the FEA model or taking into account the magnetic field-dependent or mechanical strain-related critical current density of the superconductors. The influence factors for the magnetization simulation, including the specified iterative load steps, the initial resistivity, the ramping time and the updating coefficient, are discussed in detail. The *A-V-A* formulation based IAM, implemented in ANSYS, shows its unique advantages in adjustable computation time, multi-frame restart analysis and easy-convergence.

## Keywords

---

## 1. Introduction

Magnetization current (screening current, shielding current or persistent current) effect of commercial superconductors, usually un-desired, has been extensively studied in superconducting accelerator magnets [1-7], NMR magnets and high $T_c$ superconducting coils [8-17] by using numerical simulation or experimental method. Recent studies show growing interests of using commercial FEM software, in which *E-J* power law based equations can be defined, to solve AC/DC magnetization problems for high $T_c$ superconductors. Generally the form of Maxwell's equations defined for eddy current solver in the FEM software can be $\textbf{\textit{A-V}}$ [18-22], $\textbf{\textit{T-Ω}}$ [23-26], $\textbf{\textit{T-A}}$ [27-29] or $\textbf{\textit{H}}$-formulation [30-35].

The magnetization effects can also be beneficial when we would like to trap magnetic field into high $T_c$ superconducting bulks or tape stacks [36-43]. The common techniques for magnetizing the bulk or tape stack include zero field cooling (ZFC), field cooling (FC) and pulsed field magnetization (PFM). To simulate the magnetization process the available FEM software which can solve the critical state model [44] or the flux creep model [45-46] can be FLUX2D/3D [20, 23], COMSOL [31, 33, 47], FlexPDE [48-49], GetDP [19], Photo-eddy [21-22] or ANSYS [50-51]. Among these FEM tools COMSOL shows its unique advantages in coupling user-defined partial differential equations, choosing the form of Maxwell's equations and conducting multi-physics coupled simulations [41, 52]. The other widely used multi-physics software ANSYS, available for secondary development by using ANSYS Parametric Design Language (APDL), has also been explored by scientists to solve AC magnetization problems of high $T_c$ superconductors based on the proposed Resistivity-Adaption-Algorithm [50-51, 53]. This algorithm aims at finding a final resistivity matrix for the meshed superconductor elements to fulfill the critical state model or the $\textbf{\textit{E-J}}$ power law based flux creep model. But the intermediate magnetization process and the relaxation of magnetization are missed.

This paper, for the first time, explores the magnetization process of high $T_c$ bulk superconductors during ZFC and FC magnetization by using the newly developed $\textbf{\textit{A-V-A}}$ formulation based iterative algorithm method (IAM). The IAM, implemented in ANSYS APDL, proves feasible to simulate the magnetization current of bulk superconductors in both the critical state model and the flux creep model. Specially, the $\textbf{\textit{A-V}}$ formulation is used in superconductor areas to calculate both the magnetic field and the eddy current while the $\textbf{\textit{A}}$-formulation (Ampere's Law) is used in non-superconductor areas to calculate only the magnetic field. Hence we name this method $\textbf{\textit{A-V-A}}$ formulation.

The pre-installed $\textbf{\textit{A-V}}$ formulation for eddy current solver is as follows

$$\nabla \times \left(\frac{1}{\mu}\nabla\times A\right) = -\frac{1}{\rho}\left(\frac{\partial A}{\partial t} + \nabla \cdot V\right) \qquad (1)$$

The current density $\textbf{\textit{J}}$ equals to the right side of (1). As $\nabla \cdot \textbf{\textit{J}} = 0$, we can get

$$\nabla \cdot \left[\frac{1}{\rho}\left(\frac{\partial A}{\partial t} + \nabla \cdot V\right)\right] = 0 \qquad (2)$$

For the critical state model the resistivity value can be expressed as

$$\rho = \begin{cases} 0 & if \ |\textbf{\textit{J}}| \leq J_c(B) \\ +\infty & if \ |\textbf{\textit{J}}| > J_c(B) \end{cases} \qquad (3)$$

where $J_c(B)$ refers to the magnetic field-dependent critical current density. For $\textbf{\textit{E-J}}$ power law based flux creep model the resistivity value can be expressed as

$$\rho = \frac{E_c}{J_c(B)} \cdot \left(\frac{|\textbf{\textit{J}}|}{J_c(B)}\right)^{n-1} \qquad (4)$$

where $E_c$ refers to the transitional electrical field (usually $10^{-4}$ V/m).

The computational efficiency of using the ANSYS $\textbf{\textit{A-V-A}}$ formulation based IAM is competitive with using the $\textbf{\textit{H}}$-formulation in COMSOL or other FEM tools. Other advantages of using this new method for solving the magnetization problems are concluded in this paper.

## 2. The critical state model based iterative algorithm and its application in ZFC and FC magnetization

### 2.1. ZFC magnetization - external field rises from zero to 1 T

Figure 1 (a) shows the 2D axis-symmetric half FEA model created in ANSYS for simulating the magnetization current in a disk-shaped ReBCO bulk (Φ25 mm, 10 mm thick) [54]. To approach the superconducting state an initial resistivity $\rho_0$ of $10^{-16}$ Ω·m is assigned to the ReBCO bulk for transient electromagnetic analysis.

In the electromagnetic model the 8-nodes 2D element-type of *Plane*233 is defined. *Plane*233 having the degree of freedom (DOF) of $A_Z$ is named ET-1 and *Plane*233 having the DOF of $A_Z$ and $V$ is named ET-2. Initially the ReBCO bulk is meshed with ET-2 (*Plane*233, ***A-V*** formulation) while non-superconductor areas are meshed with ET-1 (*Plane*233, ***A***-formulation, the eddy current solver is inactive when the resistivity value is not specified). After meshing the whole FEA model we apply symmetric boundary (flux normal) to the bottom line and flux parallel boundary ($A_z$=0) to the two outer air lines. The external magnetic field going through the ReBCO bulk is provided by controlling source current density of the solenoid.

In Figure 1 (b) the external magnetic field rises linearly from zero to 1 T in 500 seconds. The iterative algorithm for solving this critical state model based magnetization process is developed and plotted in Figure 1 (c) by following

a) Create 2D axis-symmetrical half electromagnetic model and divide the field ascending process into $N_1$ load steps.

b) Apply an external magnetic field of $B_0/N_1$ at "$t=T_1/N_1$" and solve this transient electromagnetic analysis as load step-1.

c) Extract the trapped current density ($J_T$) for each ReBCO bulk element after load step-1 and force $J_T$ to $J_c \cdot J_T/|J_T|$ if the bulk element is over-trapped ($|J_T|>=J_c$). This can be done by modifying their element-type to ET-1 and using the "BFE" command to force their $J_T$, meanwhile, making their eddy current solver inactive in later steps.

d) Set the external field to $2B_0/N_1$ at "$t=2T_1/N_1$" and solve this transient electromagnetic analysis as load step-2.

e) Do loop calculations for (c)-(d) by iterating $i$ through 2 to $N_1$. The final trapped $J_T$ in ReBCO bulk can be reached after solving load step-$N_1$.

Assuming the $J_c$ is constantly $3 \times 10^8$ A/m² the trapped $J_T$ in ReBCO bulk is solved and plotted in Figure 2 (a) ($B_0$=1 T, $T_1$=500 s, $N_1$=200). It can be found that all the penetrated ReBCO bulk elements carry a constant current density of $-3 \times 10^8$ A/m² at "$t$=500 s". The penetration depth in the mid-plane is 17 bulk-elements (17 x 0.25 mm). In Figure 2 (b) the trapped $J_T$ is solved by defining an *n*-value of 200 in COMSOL to approach the critical state model. The trapped $J_T$ and the penetrating depth in ReBCO bulk agree well with those shown in Figure 2 (a).

### 2.2. ZFC magnetization - external field rises from zero to 1 T and then drops to zero

In Figure 3 (a) the external field going through the ReBCO bulk rises linearly from zero to 1 T in 500 seconds and then drops linearly to zero in another 500 seconds. The iterative algorithm for solving this critical state model based magnetization process is developed and plotted in Figure 3 (b).

As described in *Section*-2.1, the element-type of the penetrated bulk elements after load step-200 has been modified to ET-1 whose eddy current solver is inactive. Thus it is necessary to reset the element-type of the penetrated bulk elements to ET-2 to continue modeling the field descending process. The penetrated bulk elements can inherit previous calculated electrical field ($\rho_0 \cdot J_T$) but the trapped $J_T$ will change polarity in the outer layer and decay in the inner layer when external magnetic field starts to decrease. A viable solution is to apply nodal voltages to the penetrated bulk elements. The applied nodal voltage for each penetrated bulk element fulfills

$$V = -2\pi \cdot r \cdot \rho_0 \cdot J_c \qquad (5)$$

where $r$ refers to the nodal radius (distance between the node and the central axis). This setting boundary can minimize the $J_T$ decay of the inner penetrated elements but will not affect the penetration of the outer layer.

Afterwards we start to model the field descending process by following

f) Set the external field to $(N_2-1)B_0/N_2$ at "$t=T_1+T_2/N_2$" and solve this transient electromagnetic analysis as load step-($N_1+1$).

g) Extract $J_T$ for each ReBCO bulk element after load step-($N_1+1$) and force $J_T$ to $J_c \cdot J_T/|J_T|$ if $|J_T|$ is larger than $J_c$.

h) Set the external field to $(N_2-2)B_0/N_2$ at "$t=T_1+2T_2/N_2$" and solve this transient electromagnetic analysis as load step-($N_1+2$).

i) Do loop calculations for (g)-(h) by iterating $j$ through 2 to $N_2$. The final trapped $J_T$ in ReBCO bulk can be reached after solving load step-($N_1+N_2$).

After load step-($N_1+N_2$) the out layer traps a constant $J_T$ of $3 \times 10^8$ A/m². But the trapped $J_T$ in the inner layer is not exactly $-3 \times 10^8$ A/m² even though we set the nodal voltage boundary at the beginning. A viable solution to approach the critical state model is to introduce an additional iterative step by using the "BFE" command to force the inner trapped $J_T$ to $-3 \times 10^8$ A/m². This will lead to the slight change of $J_T$ in the nearby bulk elements but the simulation of this iterative step obeys to the Maxwell equation.

Figure 2 (c) and (d) compares the trapped $J_T$ in ReBCO bulk through using the *A-V-A* formulation in ANSYS ($J_c=3 \times 10^8$ A/m², $B_0=1$ T, $T_1=500$ s, $T_2=500$ s, $N_1=200$, $N_2=200$) and the *H*-formulation in COMSOL. In Figure 2 (c) the ReBCO bulk traps $3 \times 10^8$ A/m² in the outer layer and $-3 \times 10^8$ A/m² in the inner layer at "$t=1000$ s". The penetration depth is 8 bulk-elements (8 x 0.25 mm) for the outer layer and 9 bulk-elements (9 x 0.25 mm) for the inner layer. In Figure 2 (d) the ReBCO bulk traps ~$2.96 \times 10^8$ A/m² in the outer layer and ~$-2.90 \times 10^8$ A/m² in the inner layer at "$t=1000$ s". The trapped $|J_T|$ is slightly lower than $J_c$ because the flux creep effect cannot be eliminated entirely even for "$n=200$". This explains why the penetration depth of the inner layer is slightly larger than that shown in Figure 2 (c).

## 2.3. FC magnetization - field drops from 1 T to zero

With regards to FC magnetization, the temperature of ReBCO bulk is kept above $T_c$ when external field rises from zero to $B_0$. This step can be realized in ANSYS-IAM by setting the element-type of ReBCO bulk to ET-1 and using the "BFE" command to force the current density to zero. Then no eddy current will be generated in the ReBCO bulk during the external magnetic field ascends.

When the ReBCO bulk becomes superconducting the external field drops linearly to zero as shown in Figure 4 (a). Details of the developed iterative algorithm for solving this magnetization process are given in Figure 4 (b).

Figure 2 (e) and (f) compares the trapped $J_T$ in ReBCO bulk through using the *A-V-A* formulation ($J_c$=3x10$^8$ A/m$^2$, $B_0$=1 T, $T_2$=500 s, $N_2$=200) and the *H*-formulation. The simulation results are similar to those shown in Figure 2 (a) and (b) but with opposite current directions.

## 3. The flux creep model based iterative algorithm and its application in ZFC and FC magnetization

### 3.1. ZFC magnetization - external field rises from zero to 1 T

Different from the FEA model shown in Figure 1 (a), the ReBCO bulk (*A-V* formulation, $\rho$ defined) and the non-superconductor areas (*A*-formulation, $\rho$ undefined) are all meshed with the element type of *Plane*13 to simulate the flux creep model.

In Figure 5 (a) the external magnetic field rises linearly from zero to 1 T in 500 seconds and holds at 1 T for another 500 seconds. The iterative algorithm for solving this flux creep model based magnetization process is developed and plotted in Figure 5 (b) which follows

a) Create 2D axis-symmetrical half electromagnetic model and divide the field ascending process into $N_1$ load steps.

b) Apply an external field of $B_0/N_1$ at "$t$=$T_1/N_1$" and solve this transient electromagnetic analysis as load step-1.

c) Extract $J_T$ for each ReBCO bulk element after load step-1.

d) Update the resistivity of each bulk element according to (6) and (7)

$$\rho^{i+1} = max\left\{\rho_0, \frac{E_c}{J_c} \cdot \left(\frac{|J_T|}{J_c}\right)^{n-1}\right\} \quad (6)$$

$$\rho^{i+1} = k_1\rho^i + (1-k_1)\rho^{i+1} \quad (7)$$

e) Set the external field to $2B_0/N_1$ at "$t$=$2T_1/N_1$" and solve this transient electromagnetic analysis as load step-2.

f) Do loop calculations for (c)-(e) by iterating $i$ through 2 to $N_1$.

g) Hold external field at zero and do loop calculations by updating the resistivity of each bulk element after each load step according to (8) and (9)

$$\rho^{j+1} = max\left\{\rho_0, \frac{E_c}{J_c} \cdot \left(\frac{|J_T|}{J_c}\right)^{n-1}\right\} \quad (8)$$

$$\rho^{j+1} = k_2\rho^j + (1-k_2)\rho^{j+1} \quad (9)$$

The specified coefficient $k_1$ and $k_2$ is used to retain partial of the old resistivity value while the coefficient "1−$k_1$" and "1−$k_2$" is used to add partial of the updated resistivity calculated by *E-J* power law. By selecting proper $k_1$ and $k_2$ we can minimize the negative effects from abrupt resistivity change during the simulation.

Figure 6 (a) and (c) plots the trapped $J_T$ in ReBCO bulk at different time steps solved by using the *A-V-A* formulation in ANSYS ($J_c$=3x10$^8$ A/m$^2$, $n$=20, $\rho_0$=10$^{-17}$ Ω·m, $B_0$=1 T, $T_1$=500 s, $T_2$=500 s, $N_1$=400, $N_2$=500, $k_1$=0.99, $k_2$=0.9). The peak $|J_T|$ drops from 2.69x10$^8$ A/m$^2$ at "$t$=500 s" to 2.21x10$^8$ A/m$^2$ at "$t$=1000 s" while the penetration depth in the mid-plane rises from 19.5 bulk-elements at "$t$=500 s" to 24.5 bulk-elements at "$t$=1000 s".

Similar simulation results can be found in Figure 6 (b) and (d) which plots the trapped $J_T$ in ReBCO

bulk by using the $H$-formulation in COMSOL. The peak trapped $|J_T|$ drops from $2.69 \times 10^8$ A/m$^2$ at "$t$=500 s" to $2.25 \times 10^8$ A/m$^2$ at "$t$=1000 s" while the penetration depth in the mid-plane rises from 20 bulk-elements at "$t$=500 s" to 24 bulk-elements at "$t$=1000 s".

### 3.2. ZFC magnetization - external field rises from zero to 1 T and then drops to zero

In Figure 7 (a) the external magnetic field rises linearly from zero to 1 T in 500 seconds, and then holds at 1 T for 500 seconds, and finally drops linearly to zero in 500 seconds. The iterative algorithm for solving this flux creep model based magnetization process is developed and plotted in Figure 7 (b). The left iterative part for the field's ascending and holding process is the same as that shown in Figure 5 (b). The right iterative part is for modeling the field descending process. Each element resistivity is updated according to (10) and (11)

$$\rho^{k+1} = max\left\{\rho_0, \frac{E_c}{J_c} \cdot \left(\frac{|J_T|}{J_c}\right)^{n-1}\right\} \qquad (10)$$

$$\rho^{k+1} = k_3 \rho^k + (1 - k_3)\rho^{k+1} \qquad (11)$$

Figure 6 (e) plots the trapped $J_T$ in ReBCO bulk at "$t$=1500 s" which is solved by using the $A$-$V$-$A$ formulation ($J_c$=3$\times 10^8$ A/m$^2$, $n$=20, $B_0$=1 T, $T_1$=500 s, $T_2$=500 s, $T_3$=500 s, $N_1$=200, $N_2$=500, $N_3$=300, $k_1$=0.99, $k_2$=0.9, $k_3$=0.99). It can be found that the penetration depth of the outer red layer rises from zero at "$t$=1000 s" to 10 bulk-elements at "$t$=1500 s" while the total penetration depth rises from 24.5 bulk-elements at "$t$=1000 s" to 25 bulk-elements at "$t$=1500 s".

Figure 6 (f) plots the trapped $J_T$ in ReBCO bulk at "$t$=1500 s" which is solved by using the $H$-formulation. The penetration depth of the outer red layer rises from zero at "$t$=1000 s" to 10 bulk-elements at "$t$=1500 s" while the total penetration depth rises from 24 bulk-elements at "$t$=1000 s" to 24.5 bulk-elements at "$t$=1500 s".

### 3.3. FC magnetization - field drops from 1 T to zero

No eddy current will be generated in the ReBCO bulk ($T > T_c$) during external field rises from zero to 1 T. When the ReBCO bulk becomes superconducting and external field descends we start to simulate the magnetization current by referring to Figure 8 (a) and (b).

Figure 9 (a)-(d) plots the trapped $J_T$ in ReBCO bulk solved by using the $A$-$V$-$A$ formulation ($J_c$=3$\times 10^8$ A/m$^2$, $n$=20, $\rho_0$=$10^{-17}$ $\Omega$·m, $B_0$=1 T, $T_2$=500 s, $T_3$=500 s, $N_2$=400, $N_3$=500, $k_1$=0.99, $k_2$=0.9) and the $H$-formulation. The simulation results are similar to those shown in Figure 6 (a)-(d) but with opposite current directions.

Figure 10 (a)-(d) plots the trapped $J_T$ in ReBCO bulk when we set the $n$-value to 40 and damp external field as shown in Figure 8 (a). In Figure 10 (a) and (c) the peak $J_T$, solved by the $A$-$V$-$A$ formulation, drops from $2.84 \times 10^8$ A/m$^2$ at "$t$=500 s" to $2.49 \times 10^8$ A/m$^2$ at "$t$=1000 s" while the penetration depth in the mid-plane rises from 18 bulk-elements at "$t$=500 s" to 21 bulk-elements at "$t$=1000 s". In Figure 10 (b) and (d) the peak $J_T$, solved by the $H$-formulation, drops from $2.84 \times 10^8$ A/m$^2$ at "$t$=500 s" to $2.55 \times 10^8$ A/m$^2$ at "$t$=1000 s" while the penetration depth in the mid-plane rises from 18 bulk-elements at "$t$=500 s" to 20.5 bulk-elements at "$t$=1000 s".

The key for modeling large $n$-value based magnetization problem in ANSYS $A$-$V$-$A$ formulation is to set more load steps or to specify tiny time steps at the beginning of magnetic relaxation.

## 4. Implement *B-H*, $J_c(B)$ and $J_c(\varepsilon)$ into the magnetization process

In this section the possibility of implementing *B-H*, $J_c(B)$ and $J_c(\varepsilon)$ into the magnetization process is investigated by "repeating" the simulation in *Section*-2.1.

### 4.1. Trapped $J_T$ in ReBCO bulk when including ferromagnetic materials

Simulation or experimental studies show that ferromagnetic materials encompassing the bulk superconductors can help to increase the trapped magnetic field in the whole system [39, 55]. It is therefore interesting to check the feasibility of introducing ferromagnetic materials into the ANSYS model. To achieve this we re-run the simulation case after modifying Air-1 in Figure 1 (a) to *B-H* curve based ferromagnetic iron.

Figure 11 (a) plots the trapped magnetic field in ReBCO bulk and ferromagnetic iron after external field rises to 1 T. The peak magnetic field reaches ~3 T in the iron. Figure 11 (b) plots the trapped $J_T$ in ReBCO bulk where we find the penetration depth is much smaller than that shown in Figure 2 (a). This is because the magnetic flux lines going through the ReBCO bulk are partially taken away by the iron. The iron shares the task with the ReBCO bulk to fight against the ascending of external magnetic field.

### 4.2. Trapped $J_T$ in ReBCO bulk when considering $J_c$-B dependence

For practical high $T_c$ superconductors the relation between $J_c$ and *B* can be described by Kim model [47, 56-57]. Here we assume the $J_c$ of ReBCO bulk material fulfills

$$J_c = J_{c0}/(1 + B/B_m) \qquad (12)$$

where $B_m$ is 1 T and $J_{c0}$ is $3 \times 10^8$ A/m$^2$.

The iterative algorithm for solving this magnetization process is developed and plotted in Figure 12. Specially, we need to extract the magnetic field *B* of each ReBCO bulk element and update the $J_c$ according to (12) after solving each load step. If the trapped $|J_T|$ for the $i^{th}$ bulk element is larger than $J_c$ we modify the element-type of the $i^{th}$ bulk element to ET-1 and use "BFE" command to force the trapped $J_T$ to $J_c \cdot J_T/|J_T|$. In addition, the bulk elements who are not penetrated are marked as "*mark* = 0" and the bulk elements who have ever been penetrated are marked as "*mark* = 1". If $|J_T|$ for the $i^{th}$ bulk element is smaller than $J_c$ but the $i^{th}$ bulk element has been penetrated before (*mark* = 1) we will use the "BFE" command to force the trapped $J_T$ to the updated $J_c \cdot J_T/|J_T|$.

Figure 13 (a) and (b) plots the magnetic field and the trapped $J_T$ in ReBCO bulk after external field rises to 1 T. The higher magnetic field region traps a lower $|J_T|$ in the whole ReBCO bulk. The relation between the trapped $|J_T|$ and *B* in each bulk element satisfies (12) seriously.

### 4.3. Trapped $J_T$ in ReBCO bulk when considering $J_c$-$\varepsilon$ dependence

Extensive studies on the relation between critical current and strain (or stress) have been carried out for commercial HTS wires or cables [58-62]. Compared to brittle Bi-2212 round wire [63-64] after heat treatment, ReBCO tape is with excellent mechanical property as the Hastelloy substrate can share the major tensile or compressive stress. This feature makes commercial ReBCO tape attractive to be used to reach a magnetic field above 20 T [65-69]. For practical ReBCO bulk materials there still lacks a scaling law for $J_c$-$\varepsilon$ under varied magnetic field. However, recent work, towards pushing the trapped magnetic field in disk-shaped ReBCO bulk, indicates there is an obvious $J_c$ reduction at high magnetic field because of the large stress level induced by Lorentz force [37]. Thus it is interesting to take into account the $J_c$-$\varepsilon$ dependence during the magnetization.

Here we will verify the feasibility of introducing the $J_c$-$\varepsilon$ dependence of the ReBCO bulk [70] into the ANSYS model by assuming the $J_c$ of ReBCO bulk material fulfills

$$J_c = J_{c0} \cdot \left[1 - \gamma \left(\frac{\varepsilon_{eq}}{\varepsilon_c}\right)^2\right] \cdot \left[\alpha + \left(\frac{1 - \alpha}{1 + \exp\left(\frac{\left|\frac{\varepsilon_{eq}}{\varepsilon_c}\right| - 1}{\beta}\right)}\right)\right] \quad (13)$$

where $J_{c0} = 3\text{x}10^8$ A/m$^2$, $\gamma = 0.1$, $\beta = 0.025$, $\alpha = 10\%$; $\varepsilon_c$ refers to the critical stain value (here it is set to $9.0\text{x}10^{-6}$) and $\varepsilon_{eq}$ refers to the Von-Mises mechanical strain.

The iterative algorithm for solving this magnetization process is developed and plotted in Figure 14. After transient electromagnetic analysis of load step-1 we perform static mechanical analysis by switching the electromagnetic element-type of *Plane*233 to the structural element-type of *Plane*183 and importing the Lorentz force data to the mechanical model. After the static mechanical analysis of load step-1 we extract the Von-Mises mechanical strain $\varepsilon_{eq}$ of each bulk element, update the $J_c$ according to (13) and switch the element-type back to *Plane*233 for electromagnetic analysis of the next load step. This iterative algorithm is quite similar to that shown in Figure 12.

Figure 15 (a)-(b) plots the Von-Mises mechanical

strain and the trapped $J_T$ in ReBCO bulk after external field rises to 1 T. The lower strain region traps higher $|J_T|$ in the whole ReBCO bulk. The relation between the trapped $|J_T|$ and $\varepsilon_{eq}$ in each bulk element satisfies (13) seriously.

All the simulations in this *Section* are carried out for the critical state model. For flux creep model we can also implement *B-H*, $J_c(B)$ and $J_c(\varepsilon)$ into the magnetization process by using similar iterative algorithm.

## 5. Discussion

To further understand the mechanism behind the ***A-V-A*** formulation based IAM we repeat the simulation case shown in Section-2.1 (critical state model, $J_c = 3\text{x}10^8$ A/m$^2$) and the simulation case in Section-3.1 (flux creep model, $J_c = 3\text{x}10^8$ A/m$^2$, $n=20$) by defining varied load steps ($N_1$), initial resistivity ($\rho_0$), ramping time ($T_1$) and updating coefficient ($k_1$). The calculation stops after external magnetic field rises to 1 T at "$t=T_1$".

### 5.1. The critical state model

#### 5.1.1. Number of specified load steps – $N_1$

Figure 16 compares the trapped $J_T$ in ReBCO bulk at "$t=T_1$" for varied simulation cases solved by critical state model. In Figure 16 (a)-(d) the ReBCO bulk is penetrated with a constant $J_T$ of -3x10$^8$ A/m$^2$ and holds the same penetrating depth when we specify different iterative load steps-$N_1$ ($\rho_0 = 10^{-16}$ Ω·m, $T_1 = 500$ s). This indicates the computation time can be shortened by using less iterative load steps without losing solution accuracy.

#### 5.1.2. Initial resistivity – $\rho_0$

Figure 16 (b) and (e)-(f) compares the trapped $J_T$ in ReBCO bulk when we assign varied $\rho_0$ to ReBCO bulk ($N_1 = 200$, $T_1 = 500$ s). Compared to Figure 16 (b) ($\rho_0 = 10^{-16}$ Ω·m) we find the same current-penetrating depth but more abnormal bulk elements nearby the penetration boundary in Figure 16 (e) ($\rho_0 = 10^{-17}$ Ω·m). In Figure 16 (f) ($\rho_0 = 10^{-15}$ Ω·m) the ReBCO bulk is not well penetrated at the "ends". This is because their trapped $J_T$ can decay easily due to the small time constant ($L/R_D$) of the bulk elements at the "ends".

### 5.1.3. Ramping time – $T_1$

Figure 16 (b) and (g)-(i) compares the trapped $J_T$ in ReBCO bulk when we specify varied ramping time-$T_1$ ($\rho_0 = 10^{-16}$ Ω·m, $N_1 = 200$). Compared to Figure 16 (b) ($T_1 = 500$ s) we find the same penetrating depth but more abnormal bulk elements nearby the penetration boundary in both Figure 16 (g)-(h) ($T_1 = 5$ s and 50 s). In Figure 16 (i) ($T_1 = 5000$ s) the ReBCO bulk is not well penetrated at the "ends" because of the low ramping rate.

When the ReBCO bulk is in superconducting state the practical resistivity is zero, far below $10^{-16}$ Ω·m. This means even a quite low ramping rate can generate a large eddy current density easily to penetrate the ReBCO bulk elements. We can re-assign a smaller $\rho_0$ to ReBCO bulk when we meet the situation in Figure 16 (f) and (i). For critical state model the trapped $J_T$ in ReBCO bulk is independent of the ramping time.

## 5.2. The flux creep model

### 5.2.1. Number of specified load steps – $N_1$

Figure 17 (a)-(d) compares the trapped $J_T$ in ReBCO bulk when we specify varied iterative load steps-$N_1$ ($\rho_0 = 10^{-17}$ Ω·m, $T_1 = 500$ s, $k_1 = 0.99$). In Figure 17 (a)-(b) the trapped $J_T$ profile of the ReBCO bulk is far from the benchmark simulation results due to the small amount of iterative load steps ($N_1 = 100$ and $N_1 = 200$). In Figure 17 (c)-(d) the ReBCO bulk traps similar $J_T$ and penetration depth which agrees well with the benchmark simulation results ($N_1 = 500$ and $N_1 = 1000$).

The resistivity change of the Element No.1250, shown in Figure 18 (a), is extracted for every time step. As shown in Figure 18 (b) the resistivity value rises from $1 \times 10^{-17}$ Ω·m at "$t=0$" to $3.87 \times 10^{-14}$ Ω·m at "$t=500$ s" when $N_1$ equals to 500 or 1000.

### 5.2.2. Initial resistivity – $\rho_0$

Figure 17 (c) and (e)-(f) compares the trapped $J_T$ in ReBCO bulk when we assign varied $\rho_0$ to ReBCO bulk ($N_1 = 200$, $T_1 = 500$ s, $k_1 = 0.99$). Compared to Figure 17 (c) ($\rho_0 = 10^{-17}$ Ω·m) we find the same trapped $J_T$ and penetrating depth in Figure 17 (e) ($\rho_0 = 10^{-16}$ Ω·m). In Figure 17 (f) ($\rho_0 = 10^{-15}$ Ω·m) the ReBCO bulk is not well penetrated at the "ends". This phenomenon is also observed in *Section* 5.1.2 and can be explained by the same reason.

Figure 18 (c) plots the resistivity change of the Element No.1250 when we set varied $\rho_0$. It can be found that the resistivity value rises from varied $\rho_0$ at "$t=0$" to $3.87 \times 10^{-14}$ Ω·m at "$t=500$ s" in all three different cases.

### 5.2.3. Ramping time – $T_1$

Figure 17 (c) and (g)-(i) compares the trapped $J_T$ in ReBCO bulk when we specify different ramping time-$T_1$ ($N_1$=500, $\rho_0 = 10^{-17}$ $\Omega \cdot m$, $k_1$=0.99). It can be found that the current-penetrating depth is larger and the peak $|J_T|$ is lower when a larger ramping time is defined. This is because large ramping time provides opportunities for the flux creep effects to relax the trapped $J_T$ [37, 47].

Figure 18 (d) plots the resistivity change of the Element No.1250 when we set varied $T_1$. The resistivity value rises from $1 \times 10^{-17}$ $\Omega \cdot m$ at "$t$=0" to $2.44 \times 10^{-12}$ $\Omega \cdot m$ at "$t$=5 s", $2.90 \times 10^{-13}$ $\Omega \cdot m$ at "$t$=50 s", $3.87 \times 10^{-14}$ $\Omega \cdot m$ at "$t$=500 s", and $4.45 \times 10^{-15}$ $\Omega \cdot m$ at "$t$=5000 s".

### 5.2.4. Updating coefficient – $k_1$

Figure 17 (c) and (j)-(l) compares the trapped $J_T$ in ReBCO bulk when we specify different updating coefficient-$k_1$ ($N_1$=500, $\rho_0 = 10^{-17}$ $\Omega \cdot m$, $T_1$=500 s). In Figure 17 (j) ($k_1$=0.9) the trapped $J_T$ profile is not far from the benchmark simulation results except for some abnormal penetrating elements. In Figure 17 (k)-(l) ($k_1$=0.8 and $k_1$=0.7) the trapped $J_T$ profile is inhomogeneous and discontinuous because the resistivity value jumps up and down and does not converge over the time. Figure 18 (e) plots the resistivity change of the Element No.1250 when we set varied $k_1$.

This section compares the simulation results of the ReBCO bulk during the field ramping process. For the field holding process it is also important to select proper $k_2$ to ensure good simulation results.

### 5.3. Computation time, advantages and disadvantages

The above simulations are conducted on a HP-Z8-G4 workstation which uses Intel(R) Xeon(R) Gold 6128 CPU @ 3.40 GHz and 3.39 GHz (two processors, each one has 6 cores and 12 threads). For critical state model based FEA model, it is meshed with 13386 elements (40555 nodes) and takes 2~3 seconds to solve each load step; for flux creep model based FEA model, it is meshed with 12767 elements (44952 nodes) and takes 2~3 seconds to solve each load step.

Compared to COMSOL or other FEM tools, the advantages of using ANSYS-IAM for magnetization simulation can be concluded as follows

a) The computation time for each load step of electromagnetic analysis is within several seconds. The total computation time is adjustable as we can choose either to achieve highly accurate simulation results (many iterative load steps) or to conduct the simulation case quickly (few iterative load steps).

b) The iterative algorithm method is based on ANSYS multi-frame restart analysis. Thus we can check the simulation results after any load steps and stop the program if we do not believe so-far result. Besides, it is feasible to restart the simulation from a solved load step. This is extremely beneficial when we have a more complex FEA model which requires a computation time of several days or weeks. In case of an accident of the workstation we can restart the simulation from where it stops.

c) The computation time can be minimized when we use the *A-V* formulation in superconductor areas and the *A*-formulation in non-superconductor areas. This can be easily done in ANSYS by selecting proper degree of freedom.

d) There is no convergence difficulty when large *n*-value is specified in the flux creep model or ferromagnetic material exists in the whole FEA model. This iterative algorithm method is advantageous when the $J_c$ is highly non-linear and influenced by multiple factors like the magnetic field, field angle, temperature and mechanical strain.

It is worth noting that the built-in form of Maxwell's equations in ANSYS can also be ***T-Ω*** formulation or ***H***-formulation through creating a new user-element [71]. ANSYS is with "open-source" secondary development environment [72].

The potential disadvantage of using the ANSYS iterative algorithm method is that the total computation time is not exactly proportional to the specified iterative steps. The program solves the first 100 iterative load steps quite quickly but solves later steps slower and slower. The solution time for above magnetization cases can be ~5 minutes for 100 load steps, ~10 minutes for 200 load steps and ~45 minutes for 500 load steps. This is not a problem for the critical state model as it can be well simulated even if we specify only 100 iterative steps, however, a lot of iterative steps (>=500) are necessary to well describe the flux creep model. This means it often takes more time to solve a flux creep model based magnetization problem.

## 6. Conclusion

A series of magnetization simulations on disk-shaped ReBCO bulk are carried out by implementing the ***A-V-A*** formulation based iterative algorithm method into ANSYS. This new method is proved feasible to simulate the ReBCO bulk's magnetization current during ZFC or FC magnetization for both critical state model and flux creep model. It is also proved feasible to include ferromagnetic materials in the whole FEA model and to include $J_c$-*B* or $J_c$-*ε* dependence of the ReBCO bulk. Good solution accuracy can be achieved if we specify enough iterative load steps and select proper initial resistivity for the ReBCO bulk during the magnetization simulation. Specially, for the flux creep model, it is quite important to define proper *k*-value which is used to retain the old resistivity. Compared to COMSOL or other FEM tools, ANSYS-IAM shows its advantages in manageable computation time, multi-frame restart analysis, easily used ***A-V-A*** formulation and easy-convergence.

## Acknowledgments

This work is supported by European Union's Horizon2020 research and innovation program under grant agreement No 777431 and European Union's Horizon2020 research and innovation program under the Marie Sklodowska-Curie grant number No. 701647. The authors would like to thank Dr. Mark Ainslie from University of Cambridge for his helpful discussions and running the benchmark simulations using the solenoid FEA model in COMSOL.

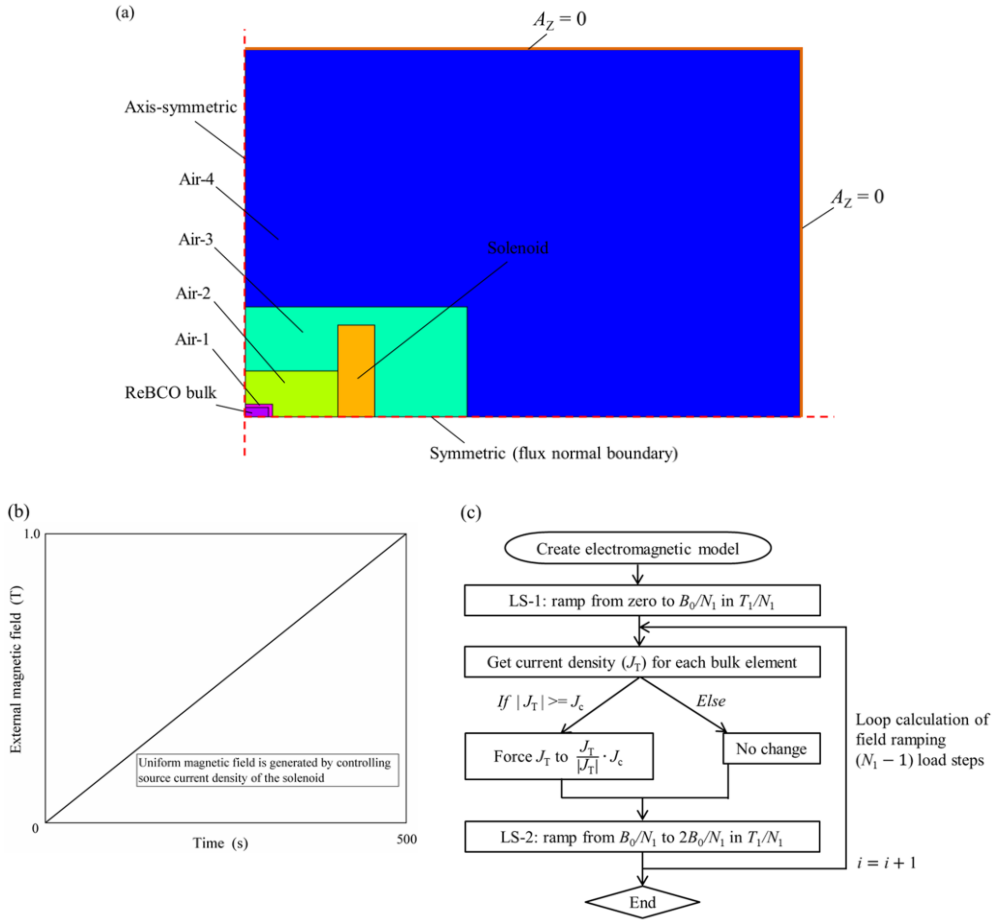

Figure 1. (a) 2D axis-symmetric half FEA model for magnetization simulation of the ReBCO bulk (the solenoid is used to generate uniform magnetic field); (b) ZFC magnetization - external magnetic field rises linearly from zero to 1 T in 500 seconds; (c) Iterative algorithm for simulating the critical state model based ZFC magnetization. This FEA model partially refers to the benchmark [54] given by COMSOL.

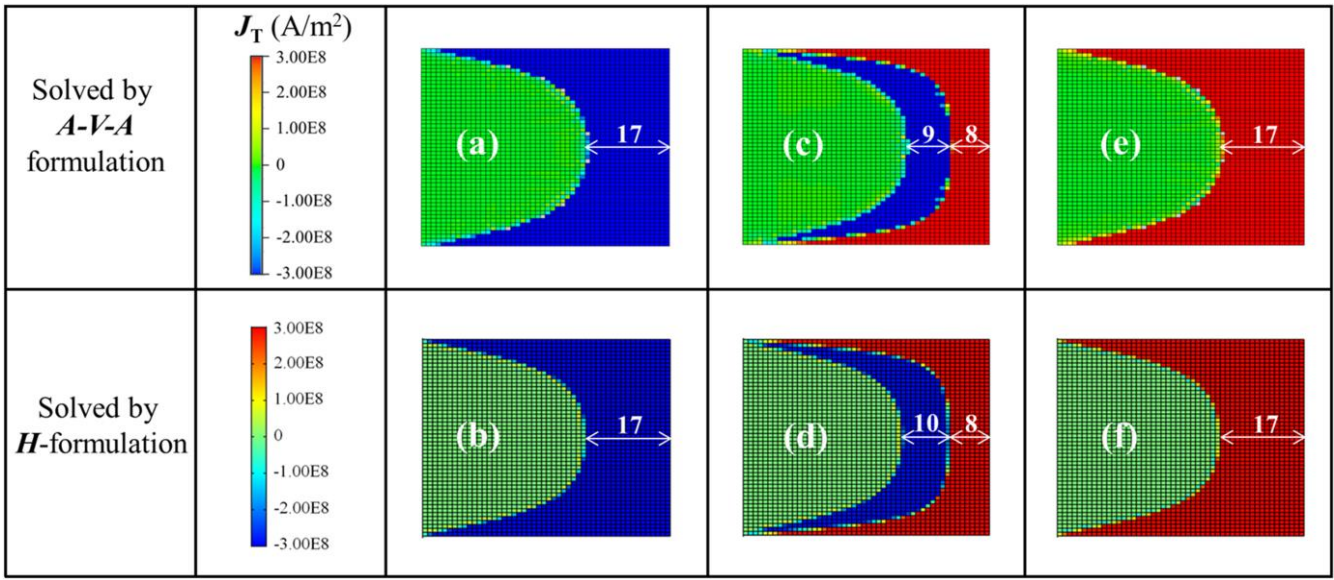

Figure 2. Trapped $J_T$ in ReBCO bulk (critical state model) after external magnetic field ascends linearly from zero to 1 T in 500 seconds by using (a) $A$-$V$-$A$ formulation and (b) $H$-formulation. Trapped $J_T$ in ReBCO bulk after external magnetic field ascends linearly from zero to 1 T in 500 seconds and then descends linearly from 1 T to zero in another 500 seconds by using (c) $A$-$V$-$A$ formulation and (d) $H$-formulation. Trapped $J_T$ in ReBCO bulk after FC magnetization from 1 T to zero in 500 seconds by using (e) $A$-$V$-$A$ formulation and (f) $H$-formulation.

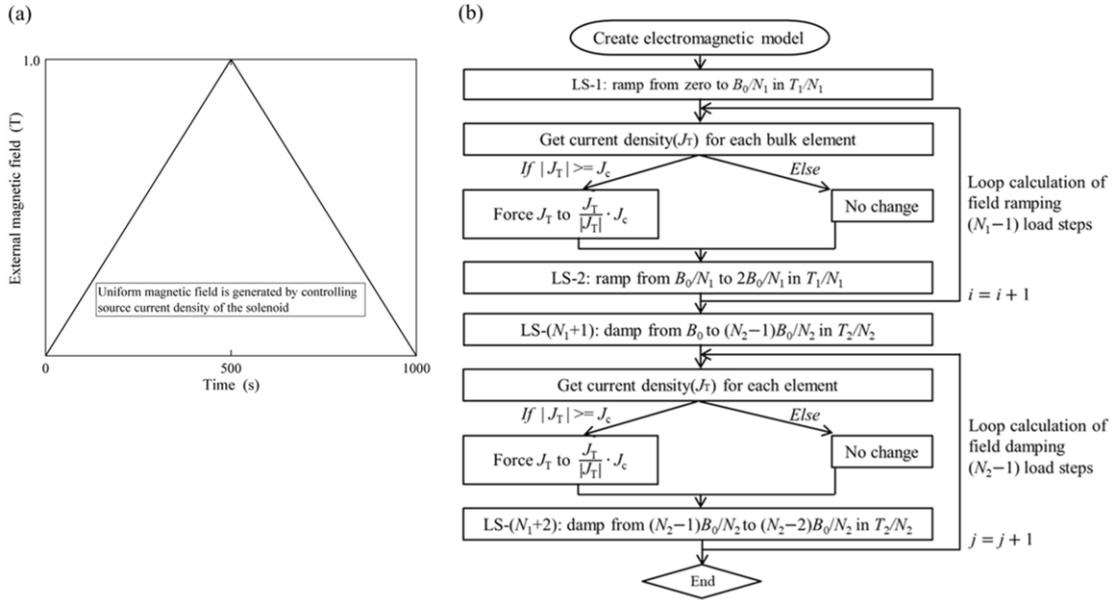

Figure 3. (a) ZFC magnetization - external magnetic field rises linearly from zero to 1 T in 500 seconds and then drops linearly to zero in another 500 seconds; (b) Iterative algorithm for simulating the critical state model based ZFC magnetization.

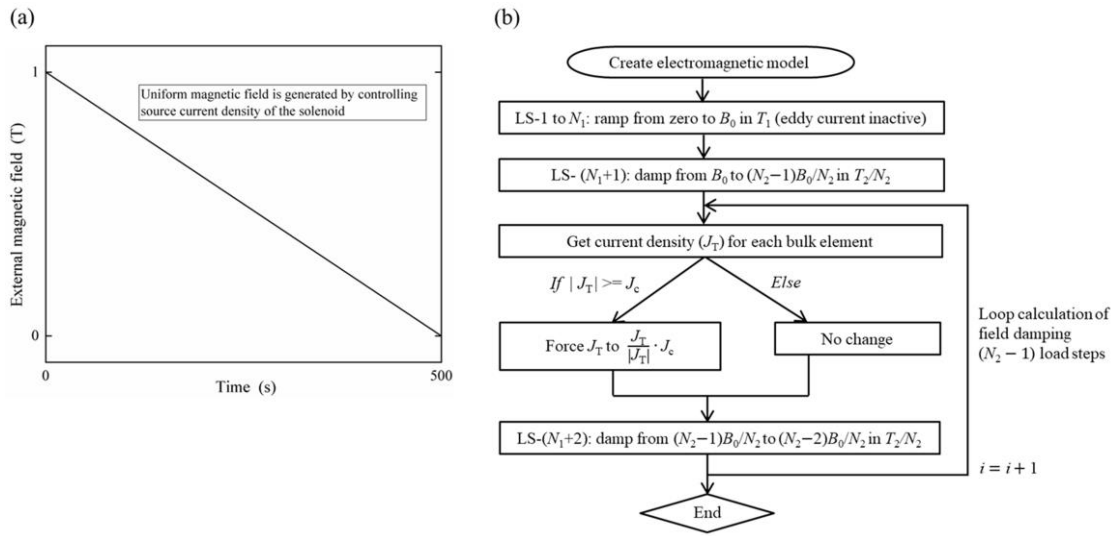

Figure 4. (a) FC magnetization - external magnetic field drops linearly from 1 T to zero in 500 seconds; (b) Iterative algorithm for simulating the critical state model based FC magnetization.

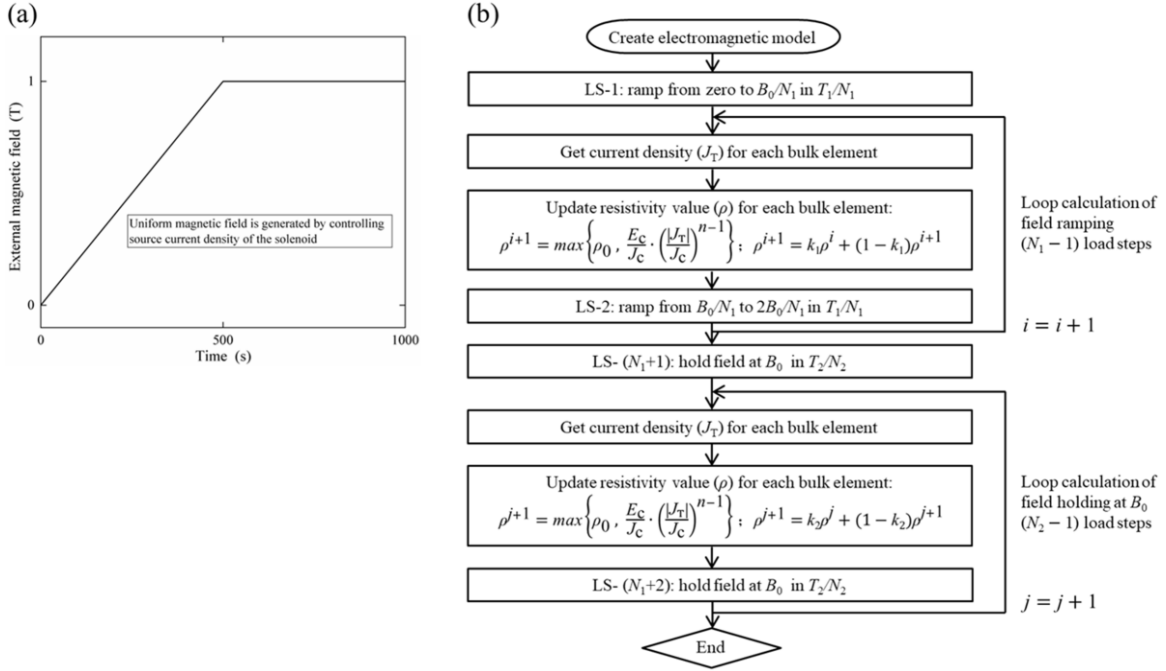

Figure 5. (a) ZFC magnetization - external magnetic field going through the ReBCO bulk rises linearly from zero to 1 T in 500 seconds and holds at 1 T for another 500 seconds; (b) Iterative algorithm for simulating flux creep model based ZFC magnetization.

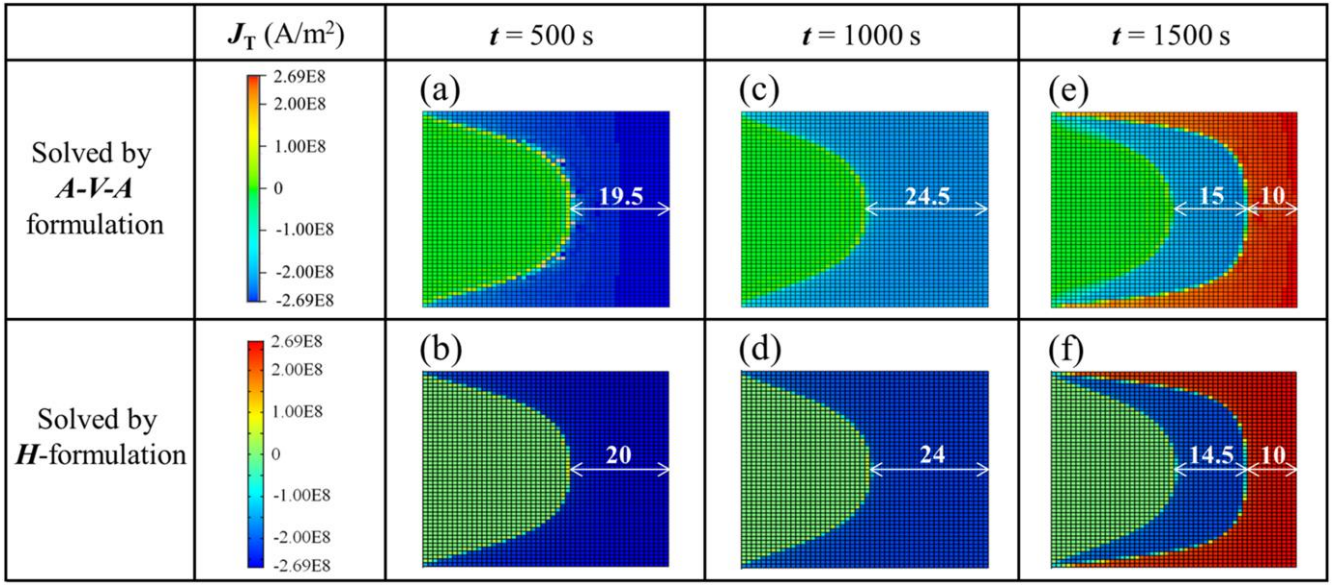

Figure 6. Trapped $J_T$ in ReBCO bulk by using the **A-V-A** formulation at (a) $t$=500 s; (c) $t$=1000 s; (e) $t$=1500 s. Trapped $J_T$ in ReBCO bulk by using the **H**-formulation at (b) $t$=500 s; (d) $t$=1000 s; (f) $t$=1500 s. The external magnetic field follows Figure 5 (a) and Figure 7 (a). ($J_c$=3x10$^8$ A/m$^2$, $n$=20)

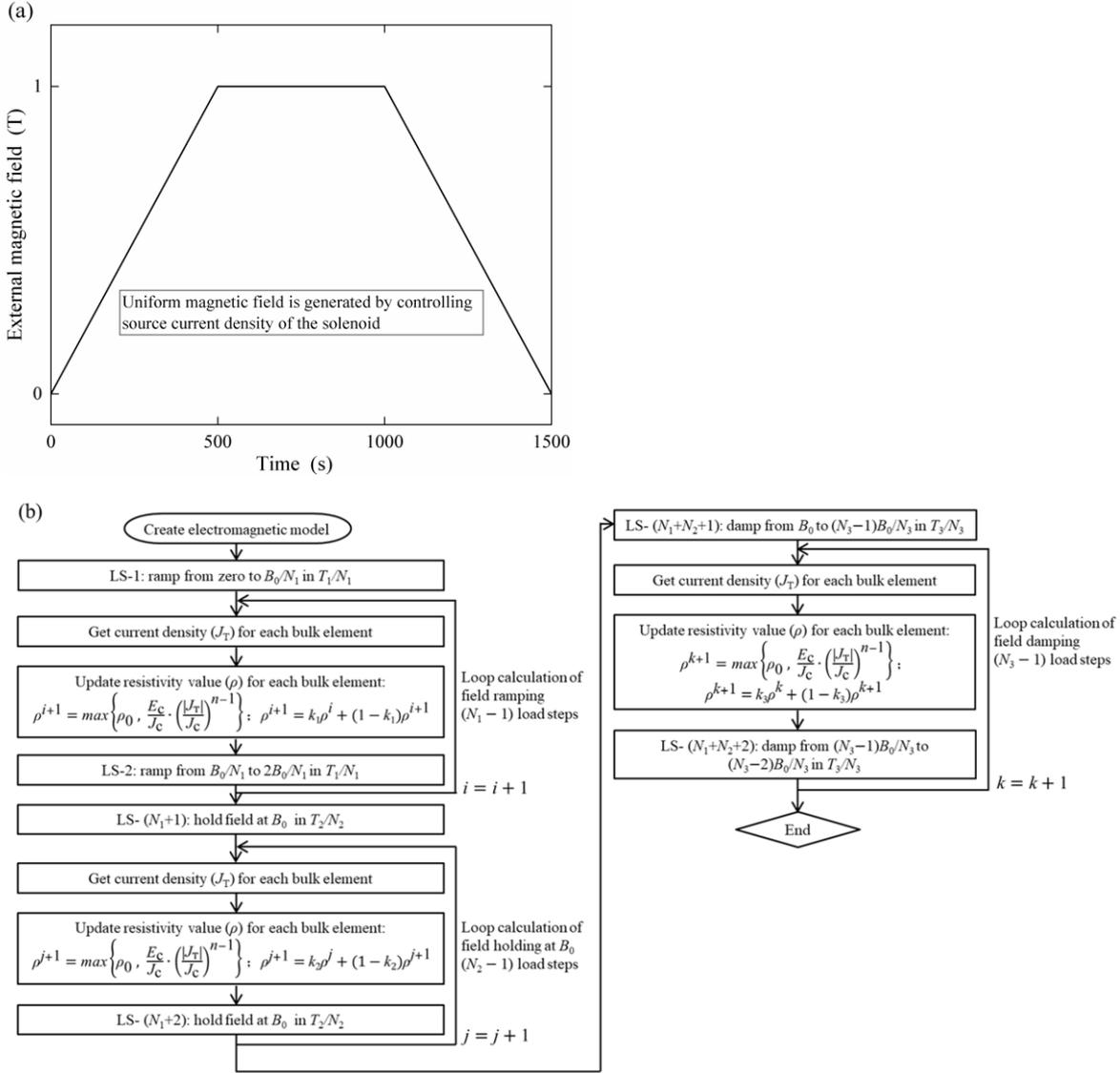

Figure 7. (a) ZFC magnetization - external magnetic field rises linearly from zero to 1 T in 500 seconds, and then holds at 1 T for 500 seconds and finally drops linearly to zero in 500 seconds; (b) Iterative algorithm for simulating flux creep model based ZFC magnetization.

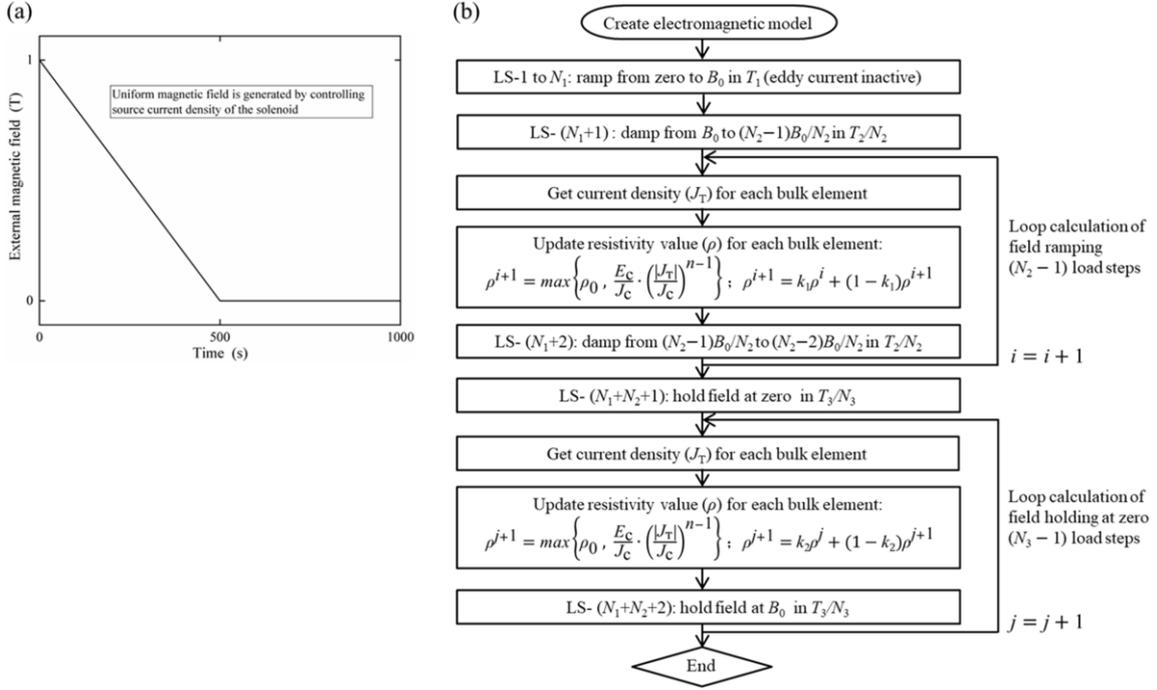

Figure 8. (a) FC magnetization - external magnetic field drops linearly from 1 T to zero in 500 seconds and holds at zero for another 500 seconds; (b) Iterative algorithm for simulating flux creep model based ZFC magnetization.

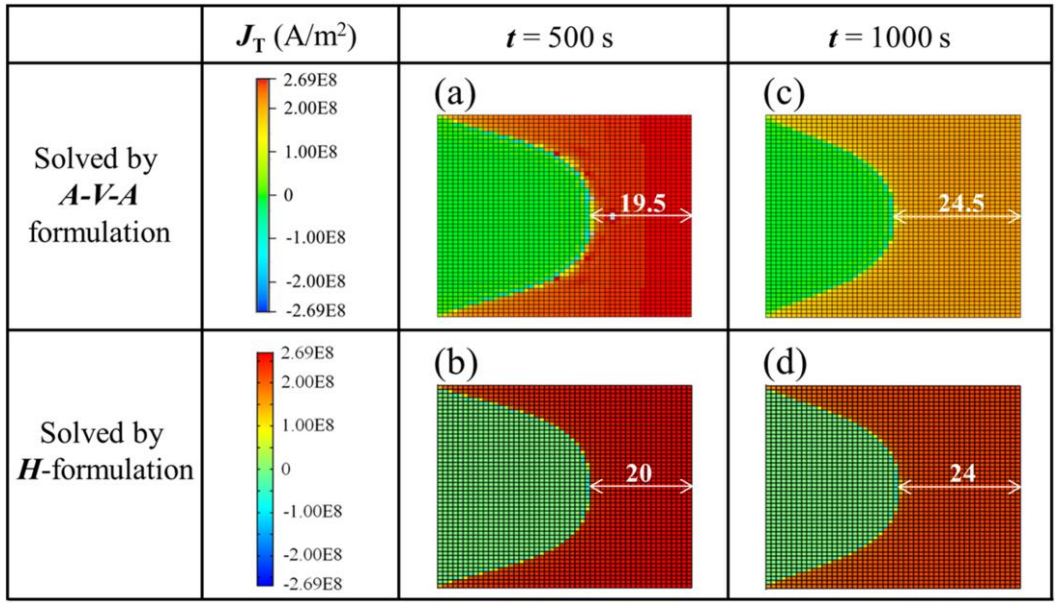

Figure 9. Trapped $J_T$ in ReBCO bulk by using the ***A-V-A*** formulation at (a) $t$=500 s; (c) $t$=1000 s. Trapped $J_T$ in ReBCO bulk by using the ***H***-formulation at (d) $t$=500 s; (e) $t$=1000 s. The external magnetic field follows Figure 8 (a). ($J_c$=3x10$^8$ A/m$^2$, $n$=20)

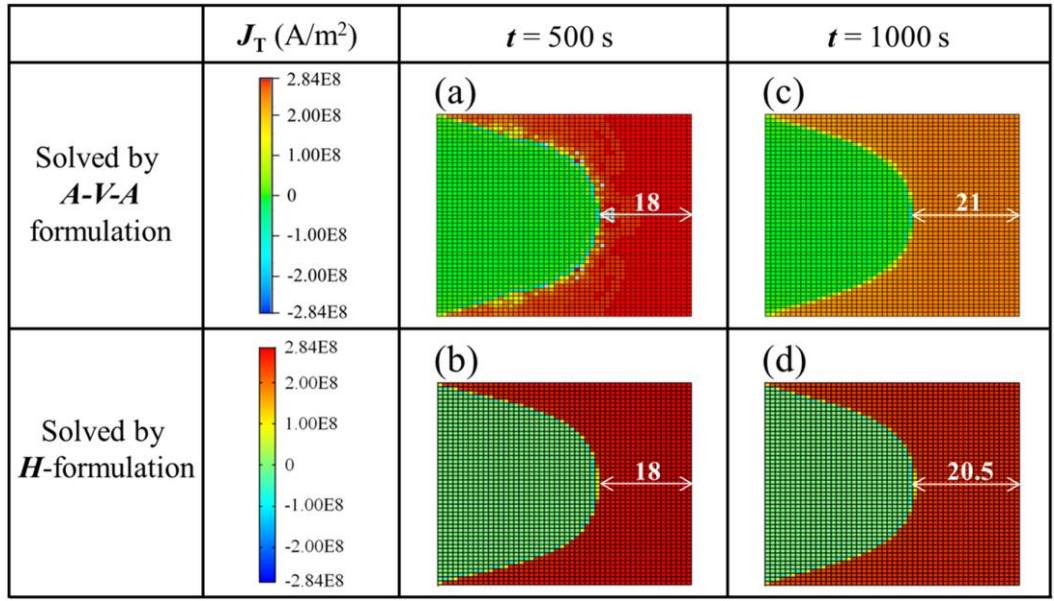

Figure 10. Trapped $J_T$ in ReBCO bulk by using the **A-V-A** formulation at (a) $t$=500 s; (c) $t$=1000 s. Trapped $J_T$ in ReBCO bulk by using the **H**-formulation at (d) $t$=500 s; (e) $t$=1000 s. The external magnetic field follows Figure 8 (a). ($J_c$=3x10$^8$ A/m$^2$, $n$=40)

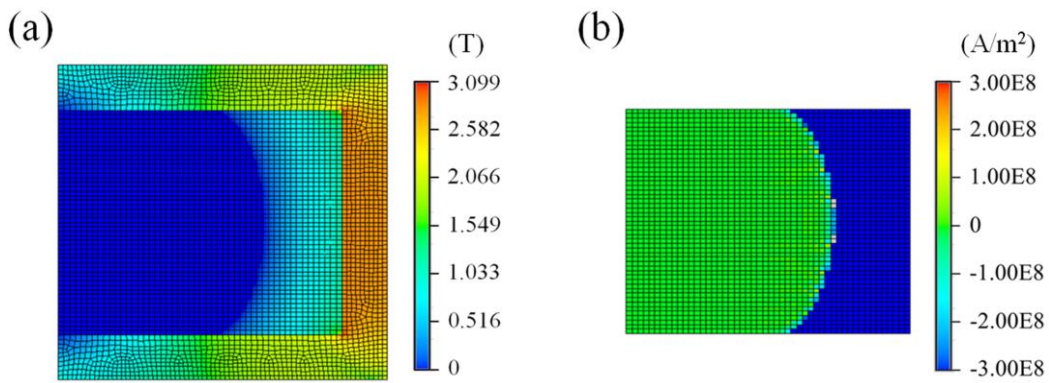

Figure 11. (a) Magnetic field map in ReBCO bulk and iron; (b) Trapped $J_T$ in ReBCO bulk. The external magnetic field rises linearly from zero to 1 T in 500 seconds (critical state model).

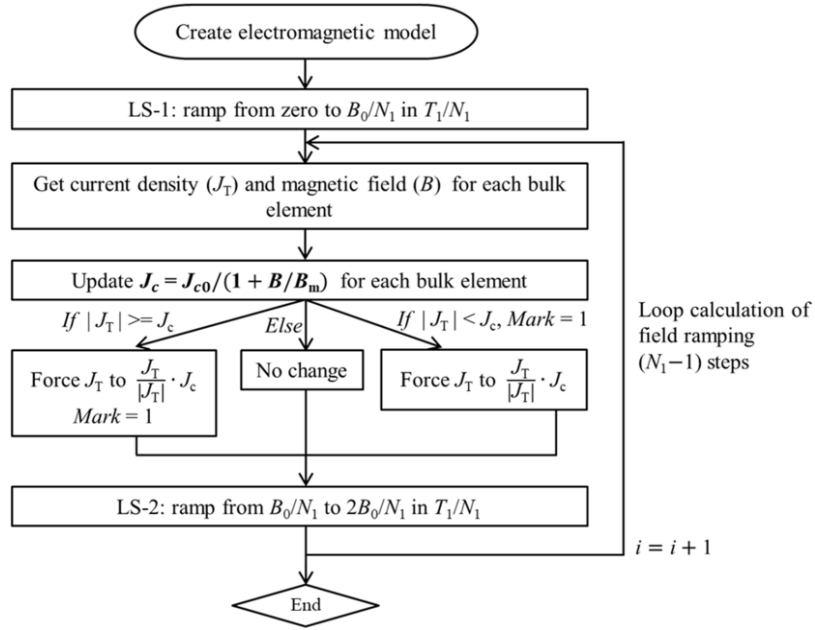

Figure 12. Iterative algorithm for magnetization simulation when considering $J_c$-$B$ dependence (critical state model).

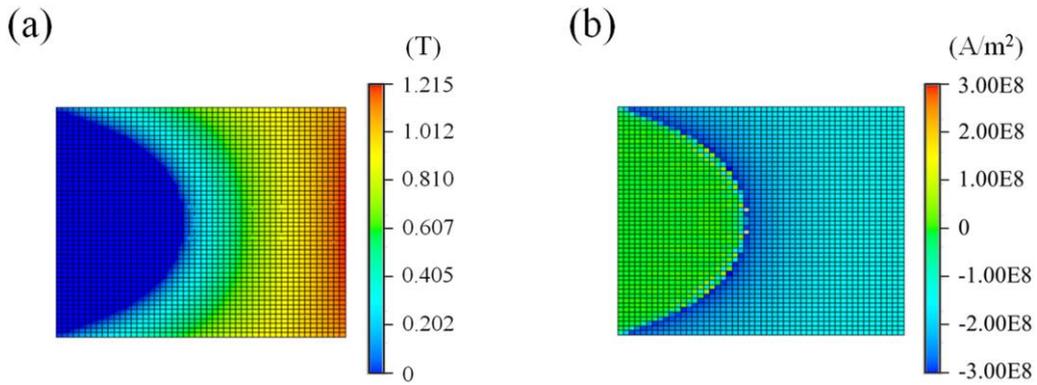

Figure 13. (a) Magnetic field map and (b) trapped $J_T$ in ReBCO bulk.

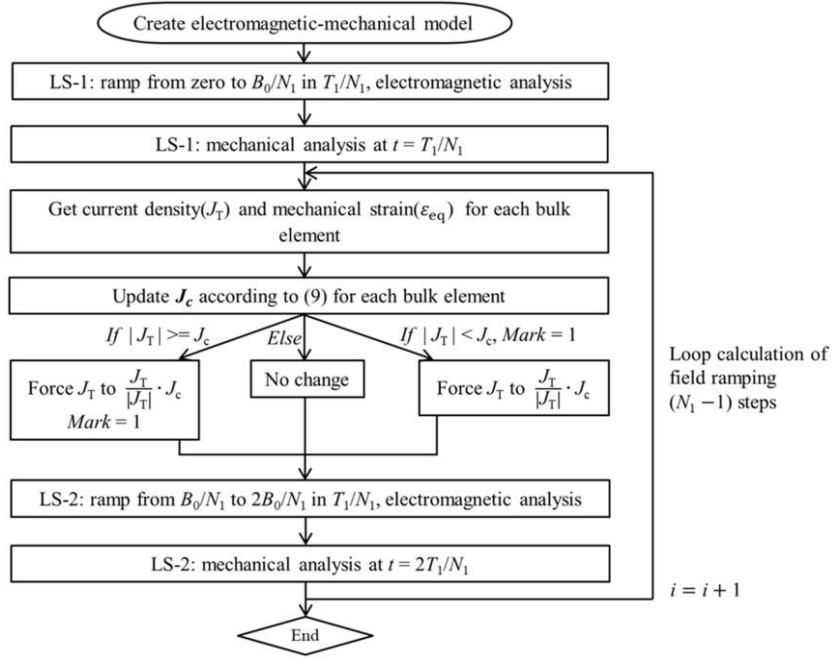

Figure 14. Iterative algorithm for magnetization simulation when considering $J_c$-$\varepsilon$ dependence (critical state model).

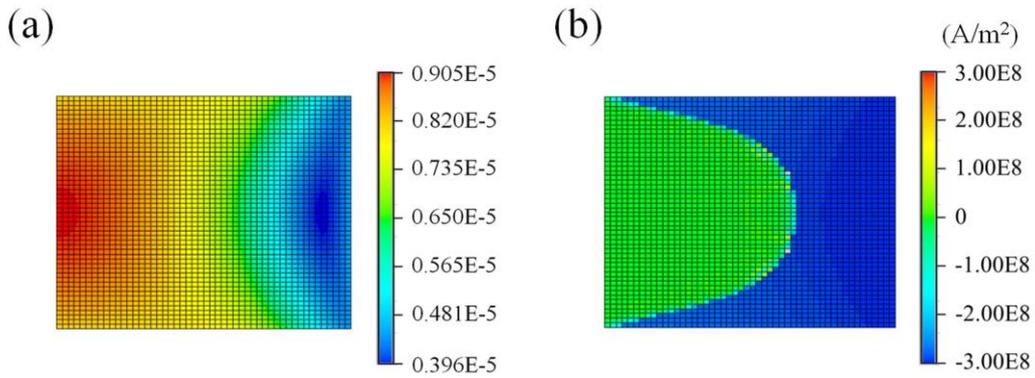

Figure 15. (a) Von-Mises mechanical strain and (b) trapped $J_T$ in ReBCO bulk.

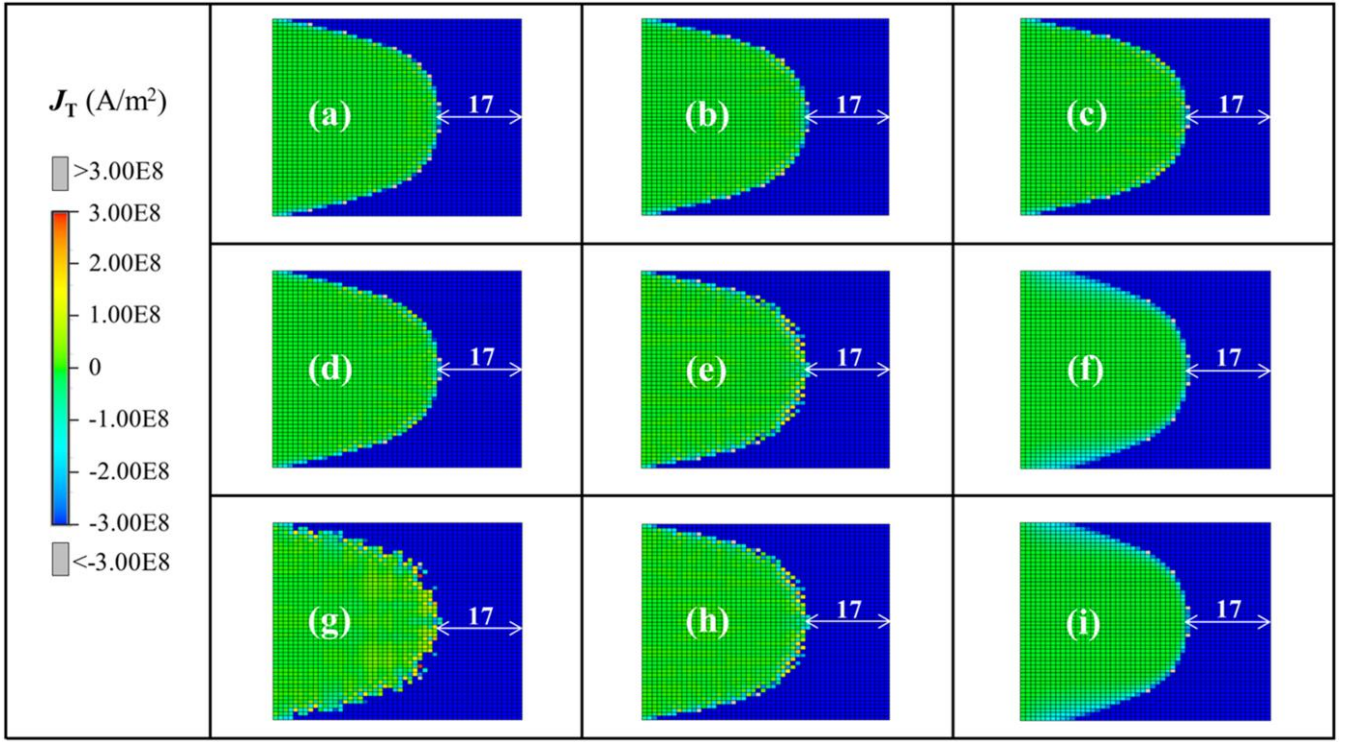

Figure 16. Trapped $J_T$ in ReBCO bulk (critical state model) after external magnetic field rises from zero to to 1 T at (a) $N_1$=100, $\rho_0 = 10^{-16}\Omega.m$, $T_1$=500 s; (b) $N_1$=200, $\rho_0 = 10^{-16}\Omega.m$, $T_1$=500 s; (c) $N_1$=500, $\rho_0 = 10^{-16}\Omega.m$, $T_1$=500 s; (d) $N_1$=1000, $\rho_0 = 10^{-16}\Omega.m$, $T_1$=500 s; (e) $N_1$=200, $\rho_0 = 10^{-17}\Omega.m$, $T_1$=500 s; (f) $N_1$=200, $\rho_0 = 10^{-15}\Omega.m$, $T_1$=500 s; (g) $N_1$=200, $\rho_0 = 10^{-16}\Omega.m$, $T_1$=5 s; (h) $N_1$=200, $\rho_0 = 10^{-16}\Omega.m$, $T_1$=50 s; (i) $N_1$=200, $\rho_0 = 10^{-16}\Omega.m$, $T_1$=5000 s.

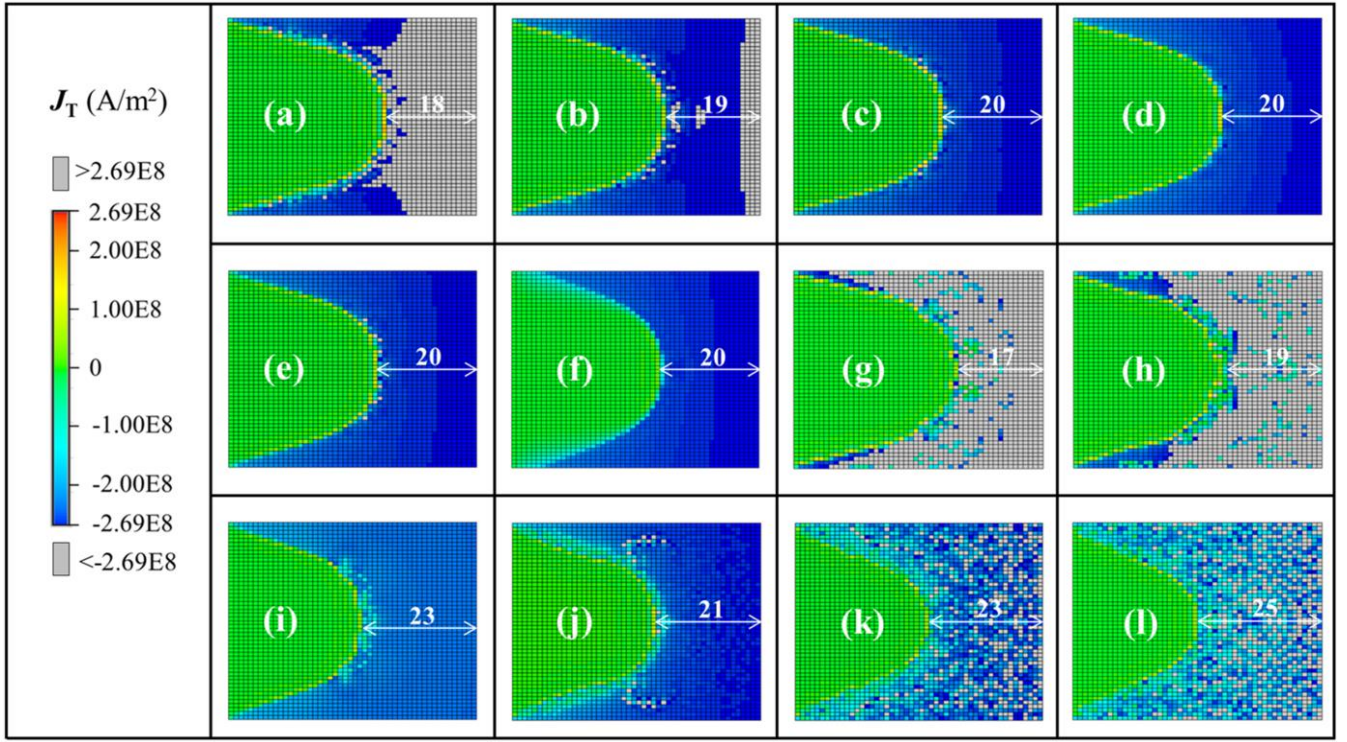

Figure 17. Trapped $J_T$ in ReBCO bulk (flux creep model, $J_c$=3x10$^8$ A/m$^2$, $n$ =20) after external magnetic field rises from zero to 1 T at (a) $N_1$=100, $\rho_0 = 10^{-17}\Omega.m$, $T_1$=500 s, $k_1$=0.99; (b) $N_1$=200, $\rho_0 = 10^{-17}\Omega.m$, $T_1$=500 s, $k_1$=0.99; (c) $N_1$=500, $\rho_0 = 10^{-17}\Omega.m$, $T_1$=500 s, $k_1$=0.99; (d) $N_1$=1000, $\rho_0 = 10^{-17}\Omega.m$, $T_1$=500 s, $k_1$=0.99; (e) $N_1$=500, $\rho_0 = 10^{-16}\Omega.m$, $T_1$=500 s, $k_1$=0.99; (f) $N_1$=500, $\rho_0 = 10^{-15}\Omega.m$, $T_1$=500 s, $k_1$=0.99; (g) $N_1$=500, $\rho_0 = 10^{-17}\Omega.m$, $T_1$=5 s, $k_1$=0.99; (h) $N_1$=500, $\rho_0 = 10^{-17}\Omega.m$, $T_1$=50 s, $k_1$=0.99; (i) $N_1$=500, $\rho_0 = 10^{-17}\Omega.m$, $T_1$=5000 s, $k_1$=0.99; (j) $N_1$=500, $\rho_0 = 10^{-17}\Omega.m$, $T_1$=500 s, $k_1$=0.9; (k) $N_1$=500, $\rho_0 = 10^{-17}\Omega.m$, $T_1$=500 s, $k_1$=0.8; (l) $N_1$=500, $\rho_0 = 10^{-17}\Omega.m$, $T_1$=500 s, $k_1$=0.7.

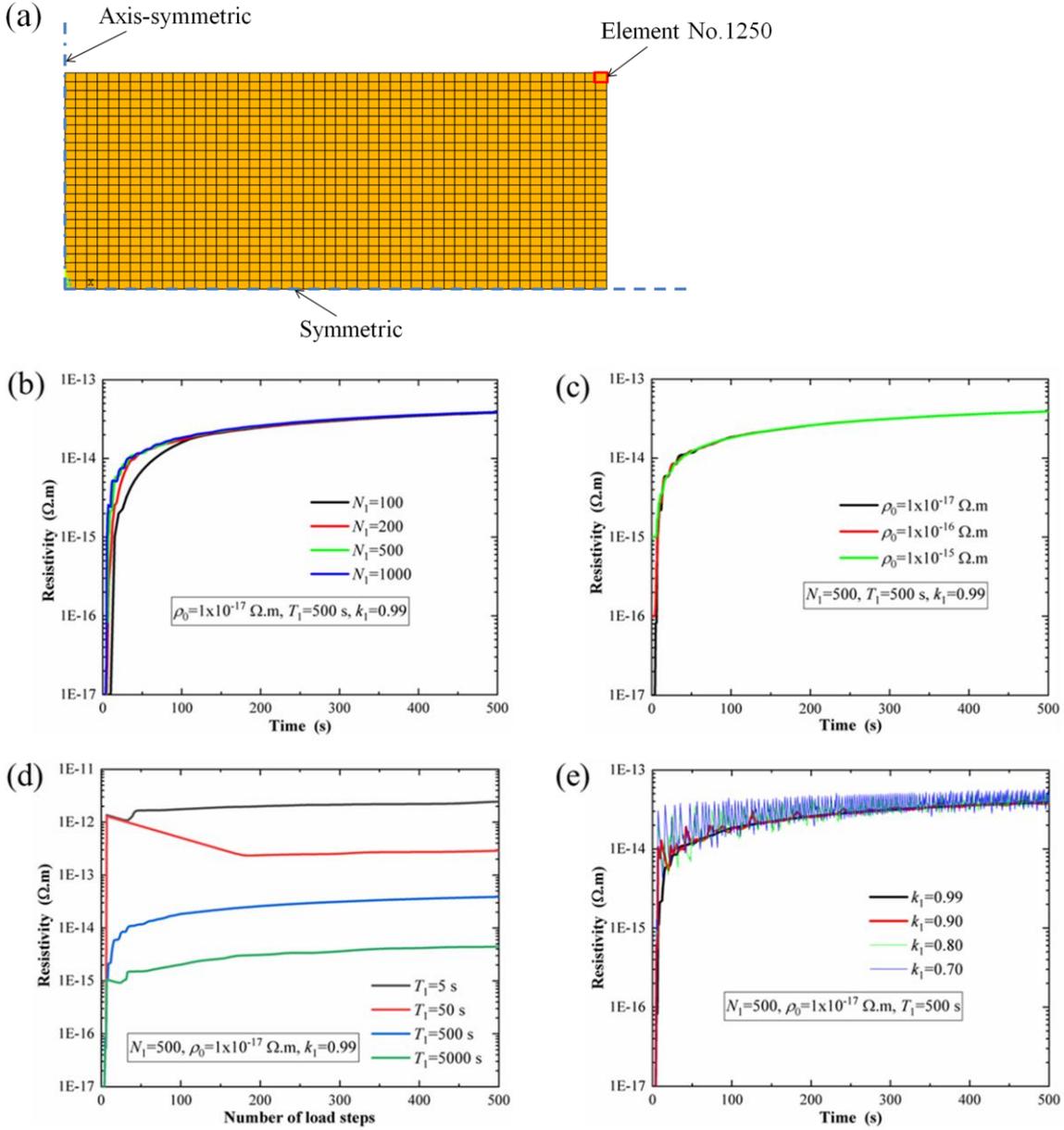